# Causal Discovery from Data Assisted by Large Language Models

Kamyar Barakati,[1,a] Alexander Molak,[2] Chris Nelson,[3] Xiaohang Zhang,[4] Ichiro Takeuchi,[4] and Sergei V. Kalinin[1, 5, b]

[1] Department of Materials Science and Engineering, University of Tennessee, Knoxville, TN 37996

[2] CausalPython.io, Warsaw, Poland

[3] Center for Nanophase Materials Sciences, Oak Ridge National Laboratory, Oak Ridge, TN 37831

[4] Department of Materials Science and Engineering, University of Maryland, College Park, MD 20742

[4] Pacific Northwest National Laboratory, Richland, WA 99354

Knowledge driven discovery of novel materials necessitates the development of the causal models for the property emergence. While in classical physical paradigm the causal relationships are deduced based on the physical principles or via experiment, rapid accumulation of observational data necessitates learning causal relationships between dissimilar aspects of materials structure and functionalities based on observations. For this, it is essential to integrate experimental data with prior domain knowledge. Here we demonstrate this approach by combining high-resolution scanning transmission electron microscopy (STEM) data with insights derived from large language models (LLMs). By fine-tuning ChatGPT on domain-specific literature, such as arXiv papers on ferroelectrics, and combining obtained information with data-driven causal discovery, we construct adjacency matrices for Directed Acyclic Graphs (DAGs) that map the causal relationships between structural, chemical, and polarization degrees of freedom in Sm-doped $BiFeO_3$ (SmBFO). This approach enables us to hypothesize how synthesis conditions influence material properties, particularly the coercive field ($E_o$), and guides experimental validation. The ultimate objective of this work is to develop a unified framework that integrates LLM-driven literature analysis with data-driven discovery, facilitating the precise engineering of ferroelectric materials by establishing clear connections between synthesis conditions and their resulting material properties.

[a] kbarakat@vols.utk.edu
[b] sergei2@utk.edu

## I. Introduction

The functionalities of solids and their structure-property relationships are governed by a complex interaction between chemical and physical degrees of freedom, which are, in turn, controlled by synthesis conditions. These effects are often interconnected through complex feedback loops that arise during the coupled dynamics of structure evolution, yielding a network of interdependencies that are critical in determining the material's final properties. Very often, processing control variables—such as temperature, pressure, and chemical environment and their process histories—differ from those that directly govern local functionalities, such as atomic-scale and extended defects, phase distributions, strain, concentration, and physical order parameter fields. Consequently, these critical local properties can only be influenced indirectly, complicating efforts to precisely control material behavior through synthesis.[1, 2]

A prime example of these complexities is found in ferroelectric materials, where chemical disorder can significantly impact polarization fields. The paradigmatic effect here will be domain nucleation, pining, field, and strain controlled dynamics of domain walls all affected by the impurities and structural defects.[3-5] However, discontinuities in the polarization field can simultaneously drive chemical segregation, leading to memory effects that cause the material to 'remember' its previous states. This bidirectional interaction between chemical and physical properties underscores the challenge of controlling materials behavior, as interventions on one degree of freedom can lead to unintended consequences in another, complicating the control and optimization of ferroelectric properties.[6, 7]

Understanding and discovering the causal mechanisms is essential for making accurate interventions and developing counterfactual strategies that can predict and manipulate material behavior. Causal models provide a pathway to go beyond the limitations of purely data-driven methods, offering insights that can guide more effective material design.[8] While experiments that involve direct interventions on specific degrees of freedom best answer *interventional* causal questions—those examining the immediate impact of one variable on another—the number of observable variables often surpasses the scope of feasible experimental controls.[9, 10] This disparity necessitates advanced approaches to causal discovery and—sometimes—addressing counterfactual questions that cannot be resolved through direct experimentation alone.[11-13]

Previously, we have explored the causal relationships in ferroelectric materials using the pairwise causal models[14] and Linear non-Gaussian Acyclic Model (LiNGAM),[15] both integrated with physics-based analyses to incorporate domain-specific insights. However, these methods face limitations when applied to complex systems, as they rely solely on observational data, and implementations used in prior work lacked the ability to incorporate prior domain knowledge effectively.

Here, we address these challenges by integrating large language models (LLMs)[16] into the causal discovery process. By fine-tuning LLMs on domain-specific literature, such as arXiv papers on ferroelectric materials, we improve the accuracy and interpretability of causal models. This approach systematically links synthesis conditions to material properties by combining literature insights with experimental data, providing a more informed framework for the design and optimization of ferroelectric materials.

## II. Data-driven causal discovery

As a model system, we selected the previously studied multiferroic material $BiFeO_3$ (BFO), modified through Sm substitution at the Bi sites and grown as an epitaxial composition-spread thin-fim. Pure BFO is characterized by its rhombohedral ferroelectric structure with polarization oriented along the $\langle 111 \rangle$ direction within the pseudocubic framework. However, substituting around 14% of Bi with Sm alters the energy landscape of the material, inducing a phase transition to an orthorhombic configuration and effectively suppressing its ferroelectric properties.[17, 18] Previously, materials within this library has been explored by scanning transmission electron microscopy (STEM) and scanning probe microscopy (SPM). These studies have been published in several previous manuscripts, and the resultant data sets were made open and are used in the present study.[19, 20]

The details of sample growth, STEM imaging preparation, and initial data analysis and quantification have been outlined in our previous publications.[21-23] The atomic positions were extracted using the DCNN based atom finder, and refined using the Gaussian peak fit.[24, 25] These analyses yield the intensities and exact position of all atomic columns corresponding to the A and B site cations in perovskite lattice of both material and substrate. Based on these, we create a set of local descriptors as defined below:

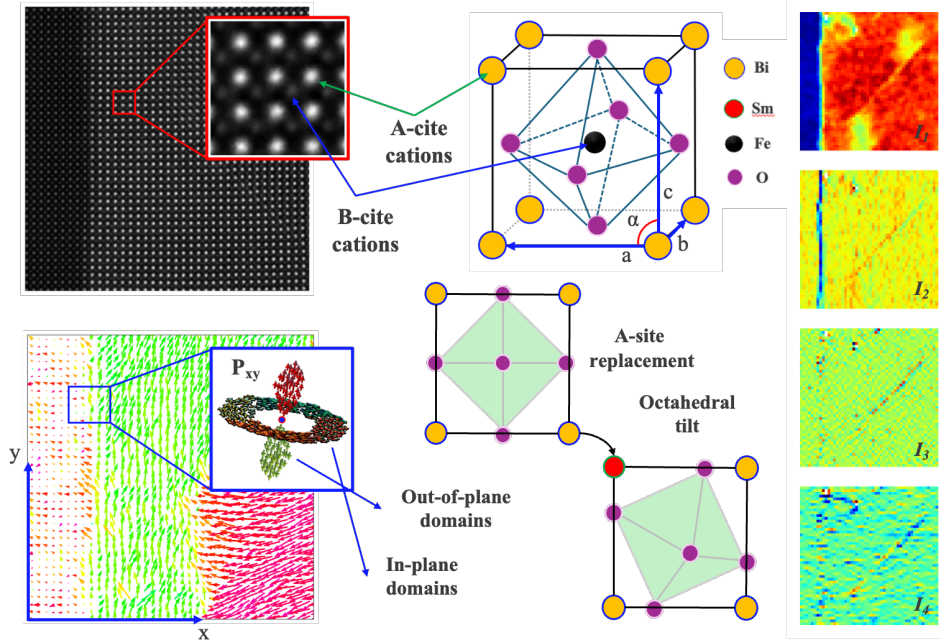

***Figure 1****: Key parameters and their visual representation for SmBFO analysis.*

***Table 1****:Key parameters and their descriptions for SmBFO analysis*

| Parameter | Description |
|---|---|
| $I_{14}$ | Intensity of A-cation columns (Bi, Sm)/ Alkali_Cations |
| $I_5$ | Intensity of B-site columns (Fe)/ Transition_Metal_Cations |
| $I_1, I_2, I_3, I_4$ | Local compositional differences and their corresponding elemental distributions |
| a | Lattice_parameter |
| c | Composition |
| α | Unit_cell_angle |
| Vol | Projected unit-cell volume/ Volume |
| $P_x$ | Electrical polarization component in the x-direction/ In_plane_polarization |

The descriptors outlined in **Table 1** and **Figure 1** provide a comprehensive foundation for analyzing the local structural, compositional, and polarization properties of Sm substituted BFO (SmBFO). Understanding the relationships between these descriptors is critical for unraveling the underlying mechanisms governing the material's behavior. For instance, variations in the intensities of A-cation and B-site columns $I_{14}$, and $I_5$ may influence structural parameters such as lattice distortions (**a**, **c**, **α**, **Vol**) and polarization component (**$P_x$**). Together, this vector (**$I_{14}$**, **$I_5$**, **a**, **c**, **α**, **Vol**, **$P_x$**) is defined for each unit cell within the field of view, enabling a detailed characterization of disorder, polarization, lattice distortions, and chemical composition at the lattice site level. This approach enables the exploration of interdependencies and mechanisms that govern the material's properties.

The PC (Peter-Clark)[26] algorithm was implemented using Gradient-Based Causal Structure Learning (gCastle)[27] as the first step in our workflow to infer Directed Acyclic Graphs (DAGs)[28] that represent causal relationships among the key descriptors of SmBFO without relying on prior domain knowledge. The process begins with skeleton identification, where every descriptor is initially assumed to be connected (**Figure 2a**). From there, tests are applied in multiple rounds to determine which connections are genuine and which can be removed. These tests vary based on the nature of the data: when values are numeric, the algorithm checks whether two descriptors remain correlated after accounting for others; when they represent discrete categories, it compares observed distributions to expected patterns; and when the underlying relationship may not follow a simple trend, kernel-based independence test is used to capture nonlinear relationships. For example, if the algorithm concludes that $I_5$ and $P_x$ are conditionally independent when controlling for the other descriptors, it removes the edge between them, producing the sparser skeleton seen in **Figure 2b.** In the second step, edge orientation is performed by detecting v-structures[29, 30]—for instance, $I_{14} \rightarrow a \leftarrow Vol$ and applying deterministic rules to orient the remaining chains, such as $I_{14} \rightarrow a \rightarrow Vol$. The final output is a DAG **(Figure 2c)**, revealing potential causal pathways among the variables and enabling further analysis of their interdependencies and directional influences.

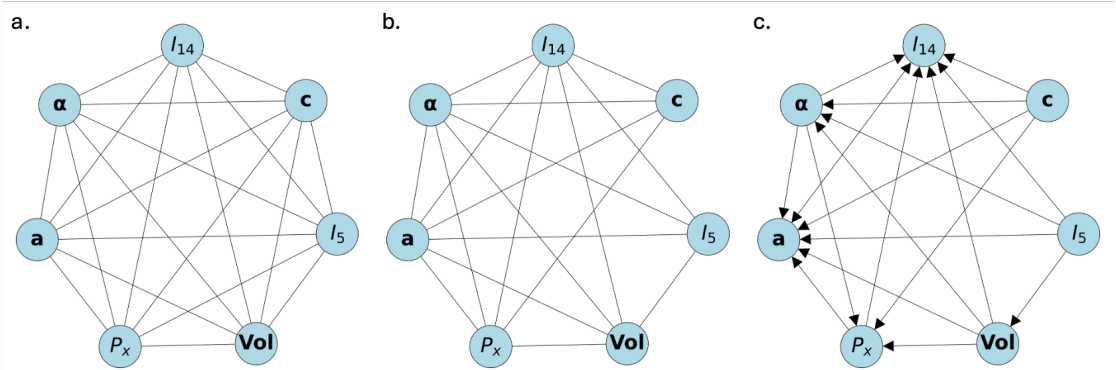

**Figure 2**: Stages of the Peter–Clark (PC) algorithm for causal discovery, where a) fully connected graph where all variables assumed to have potential dependencies, b) Skeleton graph after conditional independence tests, c) Final Directed Acyclic Graphs (DAG) with oriented edges, identifying potential causal relationships among the variables based on v-structures and logical rules.

These results provide key insights into how compositional and structural parameters interact to influence polarization and overall material properties. For example, the causal link between $I_{14}$, $\boldsymbol{a}$, and $\boldsymbol{P_x}$ suggests that modifications in A-site (through Sm substitution at the Bi sites) composition can propagate through structural changes to alter the ferroelectric properties of SmBFO. Similarly, the dependency of $\boldsymbol{Vol}$ on both $\boldsymbol{a}$ and $\boldsymbol{c}$ reflects the intrinsic coupling between lattice distortions and unit-cell geometry.

Furthermore, this causal analysis can also be performed using a sliding window approach, as demonstrated in previous studies[20]. In this technique, causal relationships are evaluated across subsets of data, providing a localized perspective on causality, which is especially useful for analyzing spatially resolved datasets or temporal sequences.

### III. Causal discovery via Large Language Models

Pure data-driven methods, such as the PC algorithm might face a number of limitations stemming from using finite sample size, presence of latent or partially latent variables, or violations of the underlying independence test assumptions. Each of these limitations might lead to smaller or larger inaccuracies in the final results, undermining the goal of the analysis. In this sense, the current generation of purely data-driven causal discovery models is not a panacea for all problems and cannot fully replace the need for domain knowledge. At the same time, incorporating domain knowledge into the process can have a protective result against the distortions stemming from some or even all the above-mentioned limitations, depending on the scenario at hand.

Limitations in problem-solving through mathematical modeling are not specific to causal discovery. Machine learning models cannot find correct solutions to ill-posed questions or those not appropriately expressed, and an in-depth understanding of the underlying chemistry is often needed to define a problem and relate it to a clear goal. Furthermore, relying solely on data-driven approaches can obscure the physical laws that govern materials' properties and behavior, hindering the understanding of fundamental materials phenomena. Data-driven analysis without physical models can be a weak lens for scientific inquiry: models can only be as good as the data they are trained on, and biases in the data can lead to inaccurate predictions. Integrating data science methods with domain knowledge can help alleviate some of these limitations.

To address these limitations, we adopted the approach proposed by A. Molak[31], which integrates prior reference data into the causal discovery process. This method allows the incorporation of domain knowledge into the DAG, refining the causal relationships suggested by data alone. Specifically, we leveraged the gCastle toolbox for causal discovery and LangChain[32] for querying large language models (LLMs).[16] Using the ChatOpenAI[33] model (GPT-4-turbo), we queried causal relationships by analyzing domain-specific literature, particularly papers available on arXiv.[34]

The querying process was designed to extract insights from the scientific corpus, evaluating causal relationships between material descriptors ($I_{14}$, $I_5$, $a$, $c$, $\alpha$, $Vol$, $P_x$) based on their scientifically recognized counterparts in **Table 1**. To ensure consistency with established conventions and prevent misinterpretation, we first queried the Web of Science database[35] to identify widely accepted terminology for these descriptors in published research. To construct a knowledge-informed causal graph, we leveraged arXiv papers on ferroelectric materials, allowing the LLM to infer relationships within a broader context rather than being restricted to a specific material system. The model as represented in **Table 2** was tasked with assessing whether a descriptor, such as $I_{14}$, influences a structural parameter like $a$, or vice versa, by interpreting causal relationships across diverse ferroelectric studies. This ensured that the extracted causal dependencies were not biased toward a single composition but rather reflected universally recognized trends in ferroelectric behavior, which were subsequently integrated into the DAG structure for further refinement.

**Table 2**: Schematic representation of the algorithmic workflow for querying causal relationships between variables using a large language model (LLM)

| ALGORITHM | |
|---|---|
| **INPUT** | Two variables (*var₁* and *var₂*) |
| **QUERY TO AGENT** | Asks if *var₁* causes *var₂*, *var₂* causes *var₁*, or neither. |
| **AGENT OUTPUT** | Provides a detailed response explaining the causal relationship (if any). |
| **LLM INTERPRETATION** | Converts the agent's response into a structured format **(0,1), (1,0), (0,0), or (-1, -1)**, where **(0,1)** indicates *var₁* causes *var₂*, **(1,0)** indicates *var₂* causes *var₁*, **(0,0)** means no causal relationship, and **(-1, -1)** means there is insufficient information to determine causality. The response should strictly follow this format without making assumptions beyond the provided data. |
| **OUTPUT** | Updates the causal graph by incorporating required and forbidden edges. |

As presented in **Figure 3 (a)**, The LLM-informed DAG has refined the causal relationships identified in the PC algorithm-derived DAG by incorporating information derived from arXiv papers. In the LLM-informed DAG, $I_{14}$ shows a direct influence on $P_x$ through **a**, providing a clearer link between A-site compositional changes and polarization. Additionally, the dependency of **Vol** on **a** and **c** is more explicitly defined, highlighting the intrinsic coupling between lattice distortions and unit-cell volume. These refinements underscore the value of integrating prior knowledge into the causal analysis to enhance the interpretability and reliability of the relationships.

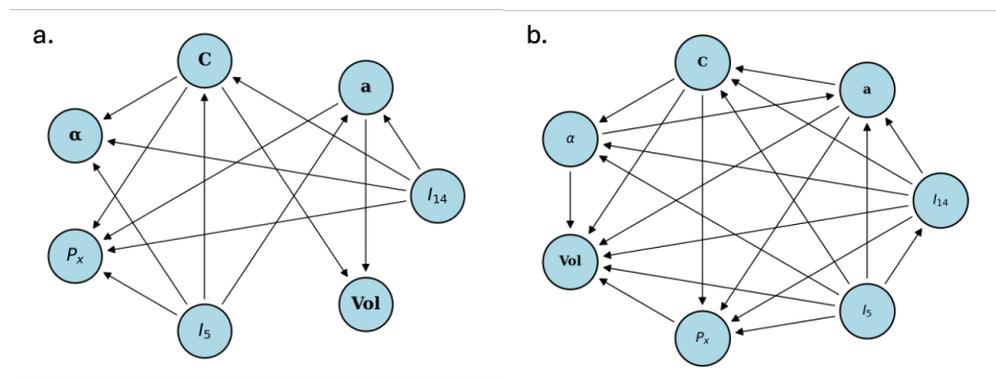

**Figure 3**: a) LLM informed DAG with oriented edges, b) PC discovered LLM-Integrated Prior Knowledge DAG with oriented edges

## IV. LLM-assisted causal discovery.

Finally, the PC algorithm was re-run, using the LLM-informed relationships as prior knowledge. This hybrid approach combines the strengths of both data-driven and knowledge-informed methods, ensuring that statistical dependencies align with domain-specific insights. The resulting DAG in **Figure 3 (b)** incorporates both the rigor of conditional independence tests and the contextual relevance of LLM-derived relationships, leading to the most reliable and interpretable structure. However, the integration of LLM-generated insights also introduces potential risks, such as biases in the training data, misinterpretation of complex relationships, or the reinforcement of spurious correlations.[36-38] Therefore, careful validation remains essential to ensure that the inferred causal pathways are both scientifically sound and meaningful.

## V. Comparative Analysis of LLM-Assisted Causal Discovery

The proposed LLM-assisted causal discovery framework presents a transformative approach for understanding complex interdependencies in materials science, particularly in systems with intricate degrees of freedom. However, it is essential to evaluate its performance and limitations by comparing it to well-established causal discovery methods, such as LiNGAM. In our prior work, LiNGAM was utilized to investigate causal relationships in ferroelectric materials, yielding significant findings that served as a foundation for this study.

From the data shown in **Figure 4**, both the LLM-Integrated Prior Knowledge DAG and the LiNGAM-discovered causal matrices reveal similar causal structures for the ferroelectric composition with x = 0. Specifically, both methods identify that the variables $I_{14}$, $a$, and $I_5$ causally influence $P_x$, while $Vol$ influences $a$. This suggests that $P_x$ is controlled by $I_{14}$, $a$, and $I_5$, $a$ is, in turn, influenced by the unit-cell $Vol$. The PC-discovered DAG integrates prior knowledge, which can enhance interpretability and capture domain-specific insights, but it may introduce bias if the prior knowledge is inaccurate. On the other hand, LiNGAM relies entirely on observed data without incorporating prior domain knowledge, making it a useful tool for uncovering causal relationships. However, its accuracy depends heavily on whether its underlying assumptions—such as linearity and non-Gaussianity—hold in the given dataset (see **Table 3** for a concise comparison of these assumptions). Violations of these assumptions, or the presence of noise or missing data, can lead to spurious causal inferences. Overall, the LLM-integrated approach is most effective when reliable prior information is available, while LiNGAM is suited for purely data-driven discovery—provided its assumptions hold. Despite their different methodologies, both methods converge on the same key pathways, underscoring the robustness of these findings for the ferroelectric system.

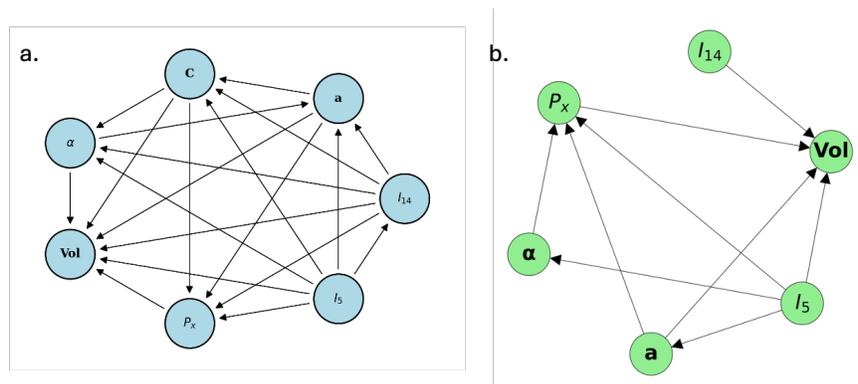

**Figure 4**: a) PC discovered LLM-Integrated Prior Knowledge DAG, b) Causal matrices LiNGAM discovered, for ferroelectric sample with fixed composition, **c = 0**. The invariant composition is excluded as a causal factor, with inferred relationships reflecting dependencies among remaining variables.

**Table 3**: Assumptions for LiNGAM and PC (Peter–Clark).

| ASUMPTIONS | LiNGAM (Original) | PC (PETER–CLARK) |
|---|---|---|
| **Functional Form** | Linear combination + non-Gaussian noise | Not strictly linear; often uses partial correlation or kernel-based tests. |
| **Noise Distribution** | Non-Gaussian errors required for identifiability | Usually assumes Gaussian errors; can use kernel-based tests for non-Gaussian data. |
| **Markov and Faithfulness** | Not explicitly invoked; relies on non-Gaussianity for identifiability | Applies Markov and Faithfulness; violations may cause spurious or missing edges.[39, 40] |
| **Absence of Unobserved Confounders** | Required (no hidden common causes) | Required (all relevant variables must be included). |
| **Sensitivity to Assumption Violations** | High if linearity or non-Gaussianity is violated | High if assumptions are violated; noise or incomplete data degrade performance. |
| **Use of Domain Knowledge** | Purely data-driven; domain expertise is optional for validation. Other variants, such as DirectLiNGAM might allow for incorporating expert knowledge.[41] | Incorporates prior knowledge (e.g., forbidden or required edges); can enhance interpretability but risk bias if inaccurate. |

**LLM-Integrated Prior Knowledge for synthesis**

By leveraging LLMs, researchers can analyze vast amounts of data from sources like arXiv to uncover causal relationships between synthesis parameters, such as temperature and pressure, and material properties like conductivity and hardness. Initial attempts suggest that without incorporating microscopic degrees of freedom, such as atomic structure and electron behavior, the findings remain too general. Combining data-driven discovery with LLM-based literature analysis allows for a comprehensive understanding of material behavior, facilitating the engineering of materials with specific functionalities. A causal graph illustrating this process would show growth conditions influencing microscopic variables, which in turn determine material properties, highlighting the importance of detailed microscopic data in the analysis.

As presented in **Figure 5**, Growth Parameters (blue nodes) represent controllable synthesis conditions such as laser energy, deposition rate, and substrate temperature, which directly influence the material fabrication process. Additional parameters like vacuum level, ambient gas pressure, and film thickness shape the physical and chemical environment, affecting critical stages of material formation. Material Properties (green nodes) in depict the resulting characteristics of the synthesized material, including polarization, structural stability, phase purity, and defects, which determine the material's performance and application potential. The integration of LLMs can significantly enhance the understanding of the causal relationships between growth parameters and material properties and provide context-aware priors for causal discovery frameworks.

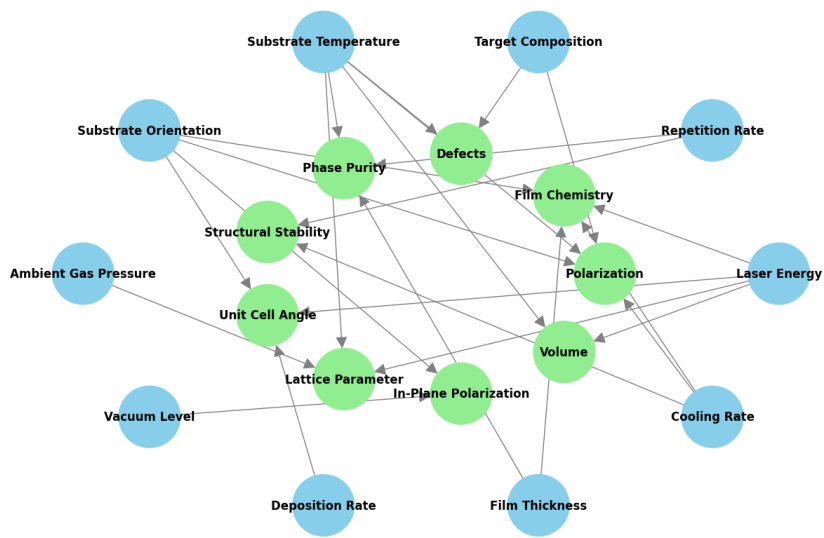

**Figure 5**: Causal Graph Depicting Relationships Between Growth Parameters and Material Properties

## Summary

In conclusion, this study demonstrates a progressive refinement in causal discovery by integrating data-driven methods with domain-specific insights. The PC algorithm without prior knowledge establishes a baseline DAG, highlighting statistical dependencies but often overlooking or misrepresenting causal pathways due to the absence of contextual understanding. The LLM-informed DAG leverages insights from published research, refining and clarifying relationships while introducing the need for validation of external knowledge. Finally, the hybrid approach—using the PC algorithm with LLM-informed prior knowledge—combines statistical rigor with domain expertise, yielding the most reliable and interpretable DAG. This integrated workflow emphasizes the importance of combining advanced machine learning techniques with domain-specific knowledge to enhance the reliability and interpretability of causal relationships, providing a powerful framework for understanding complex systems such as SmBFO and guiding future experimental efforts.

## Acknowledgements

This work (workflow development, concept) was supported (K.B., and S.V.K.) by the U.S. Department of Energy, Office of Science, Office of Basic Energy Sciences Energy Frontier Research Centers program under Award Number **DE-SC0021118**. STEM imaging (CN) was performed at the Oak Ridge National Laboratory's Center for Nanophase Materials Sciences (CNMS). The work at the University of Maryland was supported in part by the National Institute of Standards and Technology Cooperative Agreement **70NANB17H301** and the Center for Spintronic Materials in Advanced infoRmation Technologies (SMART) one of centers in nCORE, a Semiconductor Research Corporation (SRC) program sponsored by NSF and NIST. AUTHOR DECLARATIONS Conflict of Interest: The authors have no conflicts to disclose.

## Author Contributions

Kamyar Barakati: Conceptualization (equal), Data curation (lead), Formal analysis (equal), Writing – original draft (equal), Software (equal), Methodology (equal); Sergei V. Kalinin: Conceptualization (equal), Formal analysis (equal), Funding acquisition (equal), Writing – review & editing (equal), Supervision (equal); Alexander Molak: Conceptualization (equal); Methodology (equal); Chris Nelson: Data curation (equal); Investigation (equal); Xiaohang Zhang, Ichiro Takeuchi: Investigation (equal).

## DATA AVAILABILITY

The code supporting the findings of this study is publicly accessible on GitHub at [GitHub].

# References

1. A. B. Georgescu, P. Ren, C. Karpovich, E. Olivetti and J. M. Rondinelli, Machine-Learning Based Selection and Synthesis of Candidate Metal-Insulator Transition Metal Oxides, arXiv preprint arXiv:2404.08653 (2024).

2. V. Venugopal and E. Olivetti, MatKG: An autonomously generated knowledge graph in Material Science, Scientific Data 11 (1), 217 (2024).

3. L. E. C. A.K. Tagantsev, and J. Fousek, *Domains in Ferroic Crystals and Thin Films*. (Springer, New York, 2010).

4. P. Muralt, R. G. Polcawich and S. Trolier-McKinstry, Piezoelectric Thin Films for Sensors, Actuators, and Energy Harvesting, Mrs Bull 34 (9), 658-664 (2009).

5. T. M. Shaw, S. Trolier-McKinstry and P. C. McIntyre, The properties of ferroelectric films at small dimensions, Annu Rev Mater Sci 30, 263-298 (2000).

6. A. Biswas, A. N. Morozovska, M. Ziatdinov, E. A. Eliseev and S. V. Kalinin, Multi-objective Bayesian optimization of ferroelectric materials with interfacial control for memory and energy storage applications, Journal of Applied Physics 130 (20) (2021).

7. A. N. Morozovska, E. A. Eliseev, A. Biswas, H. V. Shevliakova, N. V. Morozovsky and S. V. Kalinin, Chemical control of polarization in thin strained films of a multiaxial ferroelectric: Phase diagrams and polarization rotation, Physical Review B 105 (9), 094112 (2022).

8. C. Chen, D. T. Nguyen, S. J. Lee, N. A. Baker, A. S. Karakoti, L. Lauw, C. Owen, K. T. Mueller, B. A. Bilodeau and V. Murugesan, Accelerating computational materials discovery with machine learning and cloud high-performance computing: from large-scale screening to experimental validation, Journal of the American Chemical Society 146 (29), 20009-20018 (2024).

9. T. D. Cook and D. T. Campbell, *Experimental and quasi-experimental designs for generalized causal inference*. (Figures, 2007).

10. P. Spirtes, C. Glymour and R. Scheines, *Causation, prediction, and search*. (MIT press, 2001).

11. S. Mysore, Z. Jensen, E. Kim, K. Huang, H.-S. Chang, E. Strubell, J. Flanigan, A. McCallum and E. Olivetti, The materials science procedural text corpus: Annotating materials synthesis procedures with shallow semantic structures, arXiv preprint arXiv:1905.06939 (2019).

12. X. Yu, Machine learning application in the life time of materials, arXiv preprint arXiv:1707.04826 (2017).

13. P. Veličković, G. Cucurull, A. Casanova, A. Romero, P. Lio and Y. Bengio, Graph attention networks, arXiv preprint arXiv:1710.10903 (2017).

14. M. Ziatdinov, C. T. Nelson, X. H. Zhang, R. K. Vasudevan, E. Eliseev, A. N. Morozovska, I. Takeuchi and S. V. Kalinin, Causal analysis of competing atomistic mechanisms in ferroelectric materials from high-resolution scanning transmission electron microscopy data, Npj Comput Mater 6 (1) (2020).

15. A. N. M. Chris Nelson, Maxim A. Ziatdinov, Eugene A. Eliseev, Xiaohang Zhang, Ichiro Takeuchi, Sergei V. Kalinin, Mapping causal patterns in crystalline solids, arXiv preprint arXiv:2103.01951 (2021).

16. W. X. Zhao, K. Zhou, J. Li, T. Tang, X. Wang, Y. Hou, Y. Min, B. Zhang, J. Zhang and Z. Dong, A survey of large language models, arXiv preprint arXiv:2303.18223 (2023).

17. K. Barakati, Y. Liu, C. Nelson, M. A. Ziatdinov, X. Zhang, I. Takeuchi and S. V. Kalinin, Reward driven workflows for unsupervised explainable analysis of phases and ferroic variants from atomically resolved imaging data, arXiv preprint arXiv:2411.12612 (2024).

18.      C. T. Nelson, R. K. Vasudevan, X. Zhang, M. Ziatdinov, E. A. Eliseev, I. Takeuchi, A. N. Morozovska and S. V. Kalinin, Exploring physics of ferroelectric domain walls via Bayesian analysis of atomically resolved STEM data, Nature communications 11 (1), 6361 (2020).

19.      M. Ziatdinov, C. T. Nelson, X. Zhang, R. K. Vasudevan, E. Eliseev, A. N. Morozovska, I. Takeuchi and S. V. Kalinin, Causal analysis of competing atomistic mechanisms in ferroelectric materials from high-resolution scanning transmission electron microscopy data, Npj Comput Mater 6 (1), 127 (2020).

20.      C. Nelson, A. N. Morozovska, M. A. Ziatdinov, E. A. Eliseev, X. Zhang, I. Takeuchi and S. V. Kalinin, Mapping causal patterns in crystalline solids, arXiv preprint arXiv:2103.01951 (2021).

21.      C. T. Nelson, R. K. Vasudevan, X. H. Zhang, M. Ziatdinov, E. A. Eliseev, I. Takeuchi, A. N. Morozovska and S. V. Kalinin, Exploring physics of ferroelectric domain walls via Bayesian analysis of atomically resolved STEM data, Nat. Commun. 11 (1), 12 (2020).

22.      M. Ziatdinov, C. T. Nelson, X. H. Zhang, R. K. Vasudevan, E. Eliseev, A. N. Morozovska, I. Takeuchi and S. V. Kalinin, Causal analysis of competing atomistic mechanisms in ferroelectric materials from high-resolution scanning transmission electron microscopy data, npj Comput. Mater. 6 (1), 9 (2020).

23.      M. Ziatdinov, N. Creange, X. Zhang, A. Morozovska, E. Eliseev, R. K. Vasudevan, I. Takeuchi, C. Nelson and S. V. Kalinin, Predictability as a probe of manifest and latent physics: the case of atomic scale structural, chemical, and polarization behaviors in multiferroic Sm-doped BiFeO3, Applied Physics Reviews 8 (1) (2021).

24.      A. Ghosh, M. Ziatdinov, O. Dyck, B. G. Sumpter and S. V. Kalinin, Bridging microscopy with molecular dynamics and quantum simulations: an atomAI based pipeline, Npj Comput Mater 8 (1), 11 (2022).

25.      M. Ziatdinov, A. Ghosh, C. Y. Wong and S. V. Kalinin, AtomAI framework for deep learning analysis of image and spectroscopy data in electron and scanning probe microscopy, Nature Machine Intelligence 4 (12), 1101-1112 (2022).

26.      M. Kalisch and P. Bühlman, Estimating high-dimensional directed acyclic graphs with the PC-algorithm, Journal of Machine Learning Research 8 (3) (2007).

27.      K. Zhang, S. Zhu, M. Kalander, I. Ng, J. Ye, Z. Chen and L. Pan, gcastle: A python toolbox for causal discovery, arXiv preprint arXiv:2111.15155 (2021).

28.      A. M. Lipsky and S. Greenland, Causal directed acyclic graphs, JAMA 327 (11), 1083-1084 (2022).

29.      D. Koller and N. Friedman, *Probabilistic graphical models: principles and techniques*. (MIT press, 2009).

30.      J. Pearl, *Causality*. (Cambridge university press, 2009).

31.      A. Molak, Causal, domain knowledge, https://alxndr.io/ (accessed.

32.      O. Topsakal and T. C. Akinci, presented at the International Conference on Applied Engineering and Natural Sciences, 2023 (unpublished).

33.      OpenAI, OpenAI, https://openai.com/ (accessed.

34.      C. University, arXiv, https://arxiv.org/ (accessed.

35.      clarivate, Web of Science platform, https://clarivate.com/academia-government/scientific-and-academic-research/research-discovery-and-referencing/web-of-science/ (accessed.

36.      Y. Liu, Y. Yao, J.-F. Ton, X. Zhang, R. Guo, H. Cheng, Y. Klochkov, M. F. Taufiq and H. Li, Trustworthy llms: a survey and guideline for evaluating large language models' alignment, arXiv preprint arXiv:2308.05374 (2023).

37.      J. Jiao, S. Afroogh, Y. Xu and C. Phillips, Navigating llm ethics: Advancements, challenges, and future directions, arXiv preprint arXiv:2406.18841 (2024).

38.    E. Eigner and T. Händler, Determinants of llm-assisted decision-making, arXiv preprint arXiv:2402.17385 (2024).

39.    P. Spirtes and J. Zhang, A uniformly consistent estimator of causal effects under the k-triangle-faithfulness assumption, Statistical Science, 662-678 (2014).

40.    J. Ramsey, J. Zhang and P. L. Spirtes, Adjacency-faithfulness and conservative causal inference, arXiv preprint arXiv:1206.6843 (2012).

41.    S. Shimizu, T. Inazumi, Y. Sogawa, A. Hyvarinen, Y. Kawahara, T. Washio, P. O. Hoyer, K. Bollen and P. Hoyer, DirectLiNGAM: A direct method for learning a linear non-Gaussian structural equation model, Journal of Machine Learning Research-JMLR 12 (Apr), 1225-1248 (2011).